\documentclass[%
 reprint,
superscriptaddress,
 amsmath,amssymb,
 aps,
pra,
onecolumn,
]{revtex4-2}

\usepackage[utf8x]{inputenc}
\usepackage[T1]{fontenc}
\usepackage{mathptmx} 
\usepackage{amsmath}
\usepackage{algorithm}
\usepackage{algpseudocode}
\usepackage{cancel}
\usepackage[pdftex]{graphicx} 
\usepackage{calc} 
\usepackage{enumitem} 
\usepackage{longtable}

\usepackage{color}

\usepackage{color}

\begin{document}

\title{A generative model for community types in directed networks}
 \author{Cathy Xuanchi Liu}
 \affiliation{School of Mathematics and Statistics, The University of Sydney, 2006, NSW, Sydney, Australia} 
\affiliation{Centre for Complex Systems, The University of Sydney, 2006, NSW, Sydney, Australia}
 \author{Tristram J. Alexander}
 \affiliation{School of Physics, The University of Sydney, 2006, NSW, Sydney, Australia}
 \affiliation{Centre for Complex Systems, The University of Sydney, 2006, NSW, Sydney, Australia}
 \author{Eduardo G. Altmann}
 \affiliation{School of Mathematics and Statistics, The University of Sydney, 2006, NSW, Sydney, Australia}
\affiliation{Centre for Complex Systems, The University of Sydney, 2006, NSW, Sydney, Australia}

\begin{abstract}
Large complex networks are often organized into groups or communities. 
In this paper, we introduce and investigate a generative model of network evolution that reproduces all four pairwise community types that exist in directed networks: assortative, core-periphery, disassortative, and the newly introduced source-basin type. We fix the number of nodes and the community membership of each node, allowing node connectivity to change through rewiring mechanisms that depend on the community membership of the involved nodes. 
We determine the dependence of the community relationship on the model parameters using a mean-field solution.  It reveals that a difference in the swap probabilities of the two communities is a necessary condition to obtain a core-periphery relationship and that a difference in the average in-degree of the communities is a necessary condition for a source-basin relationship.  More generally, our analysis reveals multiple possible scenarios for the transition between the different structure types, and sheds light on the mechanisms underlying the observation of the different types of communities in network data.
\end{abstract}

\maketitle

\section{Introduction}

The identification of community structures in a complex network is a fundamental step in understanding the nature of a network~\cite{GirvanPNAS2002,fortunatoPR}.  Traditional work on community structures focuses on assortative communities, in which nodes are more strongly linked to nodes in the same community than to nodes outside the community~\cite{GirvanPNAS2002,fortunatoPR,newman2006modularity,Blondel_2008}. Another type of network structure is core-periphery~\cite{coreperiphery1,coreperiphery2,peter,zhang2015identification,barucca2016disentangling,kojaku2018core}, consisting of a strongly inter-connected group of nodes (the core) influencing a sparsely connected group of nodes (the periphery).  These two examples rely on a pairwise classification approach~\cite{betzel2018non} that determines the type of community structure based on the relative strength of connections within a community compared to connections to a second community.  While this approach can uncover both assortative and core-periphery community structures~\cite{kojaku2018core,betzel2018non,urena2023assortative}, it can also reveal new types of community structure~\cite{liu2023nonassortative}.

The type of pairwise community structure for two communities $r$ and $s$ is determined by the density $\omega_{rs}$ of edges between nodes in these communities, defined as
\begin{equation}\label{eq.omegaU}
\omega _{rs} =
\begin{cases}
    \sum_{i\in r , j \in s} \frac{A_{ij}}{N_{r}N_{s}},& \text{if } r \neq s\\
    \sum_{i\in r , j \in s} \frac{A_{ij}}{N_{r}(N_{s}-1)},  & \text{if } r = s
\end{cases}
\end{equation}
where $A_{ij} \in \{0,1\}$ with $A_{ii}=0$ corresponds to the adjacency matrix of a directed graph with no self edges. For undirected networks $A_{ij}=A_{ji}$, three types of pairwise interactions are defined~\cite{kojaku2018core,betzel2018non,urena2023assortative} depending on the ranking of the corresponding density matrix: assortative, disassortative, and core-periphery. 

\begin{figure*}[h]
\includegraphics[width=0.85\textwidth]{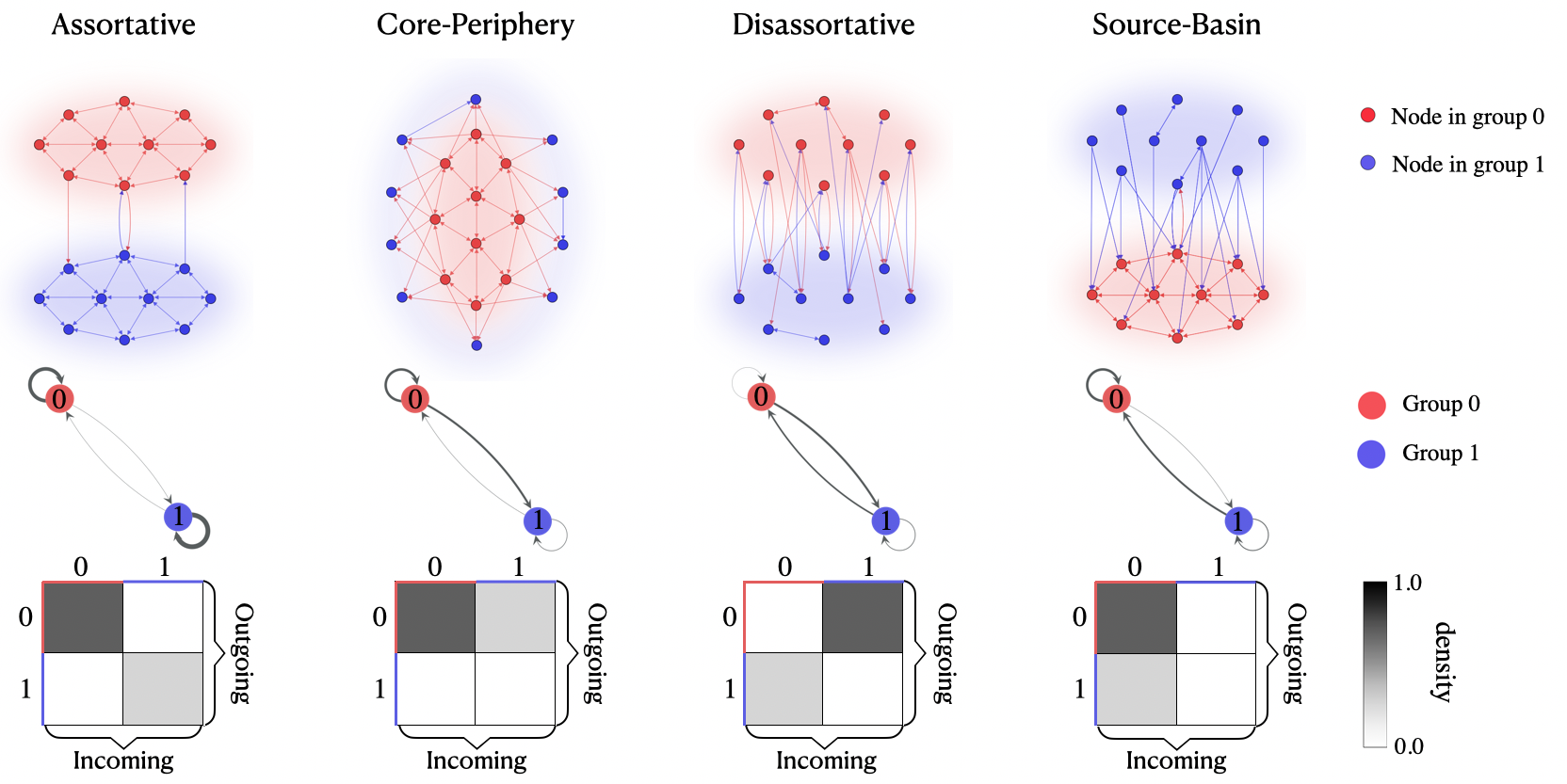}
\centering
\caption{Four types of communities in directed graphs with two groups $g$, red (0) and blue (1). (Top) Representative network showing the community-structure type. (Middle) Block Network in which each node represents one group. (Bottom) Density matrix $\omega_{rs}$ defined in Eq.~(\ref{eq.omegaU}), with the two largest matrix entries coloured in gray (see legend).. The classification of the four types is given in Eq.~(\ref{eq.inequalities}), as proposed in Ref.~\cite{liu2023nonassortative}.}\label{fig.1}
\end{figure*}

In this work we are interested in the possible structure types in {\it directed} networks (i.e., when $A_{ij} \neq A_{ji}$ is possible).  We consider the edge direction to represent the flow of information, i.e., $A_{ij}=1$ if node $i$ influences node $j$. In this case, the pairwise (with $r,s \in \{0,1\}$) community types can be classified according to the ranking of four density values ${\bf \omega} =\{\omega_{00},\omega_{01},\omega_{10},\omega_{11}\}$. In our previous work~\cite{liu2023nonassortative}, we introduced a classification of the possible pairwise structures in directed networks into four types: assortative (A), core-periphery (CP), disassortative (D), and source-basin (SB), as illustrated in Fig.~\ref{fig.1}. Mathematically, they are defined by grouping the $12$ possible rankings of $\omega_{rr},\omega_{rs},\omega_{sr},$ and $\omega_{ss}$ (with $r\neq s$) into four types as follows:

\begin{equation}
\begin{cases}\label{eq.inequalities}
    \text{Assortative (A)} ,& \text{if } \max{(\omega_{rs},\omega_{sr})} < \min{(\omega_{rr},\omega_{ss})}, \\
    \text{Core-Periphery (CP)} ,& \text{if } \min{(\omega_{rr},\omega_{rs})} > \max{(\omega_{sr},\omega_{ss})},\\
    \text{Disassortative (D)} ,& \text{if } \min{(\omega_{rs},\omega_{sr})} > \max{(\omega_{rr},\omega_{ss})}, \\
    \text{Source-Basin (SB)},& \text{if } \min{(\omega_{rr},\omega_{sr})} > \max{(\omega_{rs},\omega_{ss})}.
\end{cases}
\end{equation}
The key difference between the known CP and the new SB structure type is that the flow of information in the CP is predominantly from the  group with high internal connections (core) to the one with weak internal connections (periphery) ~\cite{van2014finding,villeseche2023presence} while in the SB case it follows from the group with weakly internal connections (source) to the one with high internal connections (basin). Our analysis of empirical networks~\cite{liu2023nonassortative} found that while assortative structures are dominant, all types appear in complex networks. The mechanism leading to these structures remain elusive, particularly in the source-basin case. 

Simple network-evolution models have been fundamental to gaining a deeper understanding of the origin of network properties~\cite{newman2003,gross_adaptive_2009}. Network growth models producing community structures which are assortative~\cite{small2008} and disassortative~\cite{klemm2002} have been introduced, providing insights into the drivers of these community types, while core-periphery community structures have been found to emerge when a small amount of assortative attachment occurs in the presence of preferential attachment~\cite{urena2023assortative}.  Rewiring processes applied to existing networks have shown how assortative community structures may emerge from random networks~\cite{xulvi-brunet2004}.  More broadly, the drivers of segregation have been uncovered in Schelling-type network models~\cite{henry2011}. While most studies focused on undirected networks, models with edge direction that combine preferential attachment and homophily have been proposed and used to study the inequality in network algorithms~\cite{espin-noboa2022}.  In line with this tradition, in our work we introduce a simple model that can explain the emergence of all four community structure types observed in directed networks. 

Our model considers generic processes that change the connectivity of nodes -- adding, removing, or rewiring edges -- and we observe the structure types that emerge as the long-term consequence of this dynamical process of network evolution. Our main findings are the parameters and conditions that lead to each of the community types. For instance, we show that performing only rewiring moves, it is possible to obtain A, D, and CP but not SB types of community structures and that an asymmetry in the rewiring mechanisms between the groups is a necessary condition to obtain CP. We also show how the SB community structure requires a variation in either the average degree of the groups or on their size (in number of nodes) and that in these cases SB appears when edges are added or removed randomly at high enough rates.

\section{Model}

\subsection{General setting}
Our goal is to find a simple network-evolution model that reproduces all four types of community-structure discussed in the Introduction. Accordingly, we consider the following simplifying assumptions:

\begin{itemize}
\item[(i)] the number of nodes is fixed, $N(t)=N=$ constant, and we focus on the evolution of the network connectivity $A_{ij}(t)$ over time $t$.
    \item[(ii)] the allocation of each node in one of the two groups is known and remains unchanged over $t$.  We denote by $g(i)$ the group $g\in\{0,1\}$ of node $i$. The sizes of each group (the number of nodes belonging to them) is given by $N_g = \sum_{i=1}^N \delta(g(i)-g)$, where $\delta(x)=1$ if $x=0$ and $\delta(x)=0$ for $x\neq0$. 
    \item[(iii)] nodes in the same group $g$ are identical (i.e., follow the same evolution rules). 
    \item[(iv)] nodes have agency in determining their incoming edges but not their outgoing ones (i.e., they can deliberately choose who they are influenced by but not who they influence).
\end{itemize}

We are interested in the equilibrium properties ($t\rightarrow \infty$) of large ($N\rightarrow \infty$) sparse ($\langle A_{ij} \rangle \rightarrow 0$) networks. 
 The choice of the initial network and group allocation is essential to determine the sizes $N_g$ of each group, as discussed above, but otherwise it will play no significant role in the final state of the network or the community-relationship between the groups (except in particular cases in which $\langle z \rangle$ remains constant). In our simulations we start with an Erdős-Rényi directed graph with $N$ nodes, $\frac{\langle z\rangle}{N-1}$ edge creation probability, and random group allocation of nodes. 

\subsection{Network evolution}\label{ssec.evoloution}

We will evolve the connectivity of the network over time $t$ step by step i.e. $A_{ij}(t) \mapsto A_{ij}(t+\Delta t)$; compute the density matrix in Eq.~(\ref{eq.omegaU}) at each time iteration $t\mapsto t+\Delta t$; and focus on the community-type classification defined in Eq.~(\ref{eq.inequalities}). To evolve $A(t)$, we randomly pick a {\it focal} node $i$ and decide at random between two types of modification of its incoming edges $A_{ji}$ (moves): a swap (with probability $P^S$) or a change in number (with probability $1-P^S$) defined as follows:

\paragraph{Swap move:}
In this move we intend to model the extent into which nodes prefer assortative edges (i.e., edges from nodes in the same group). This is implemented by randomly picking two edges 
\begin{itemize}
    \item one existing (incoming) edge $k\mapsto i$ of node $i$ (such that $A_{ki}=1$); and 
    \item a non-edge $j\mapsto i$ (i.e., $A_{ji}=0$) as a candidate edge,
\end{itemize}
 and deciding whether we rewire from edge $k\mapsto i$  to edge $j\mapsto i$ (at time $t+\Delta t$) depending on the consistency between their groups $g(k),g(j),$ and $g(i)$. We give a preference to an assortative edge (i.e., one coming from a node in group equal to $g(i)$) with probability $P^A=P^A_g=P^A(g(i))$, in which case we keep the edge coming from a node with group equal to $g(i)$. This implies that with probability $1-P^A$ a preference is given to disassortative edges, in which case we keep the edge coming from a node with group different from $g(i)$. If both candidate edge $j\mapsto i$ and existing edge $k \mapsto i$ are such that $g(j)=g(k)$, we randomly pick one to keep and delete the other. We investigate the simplest case in which the two edges are chosen randomly (from all existing and non-existing edges). Alternative choices could consider local searches. 

\paragraph{Change move:} In this move we intend to model the process of adding and removing edges, and therefore how selective nodes are about the nodes they are influenced by. We first decide either to remove or add edges with probability $\frac{1}{2}$ (as shown in Appendix~\ref{app.evolution}, this choice is taken without loss of generality). To add an edge, we pick a random node $j$ that does not have an edge pointing to $i$ (i.e., $A_{ji}=0$) and add this edge $j\mapsto i$. To remove, we delete with probability $\alpha=\alpha_g=\alpha(g(i))$ each of the incoming edges of node $i$.

\begin{figure*}[h]
\includegraphics[width=0.7\textwidth]{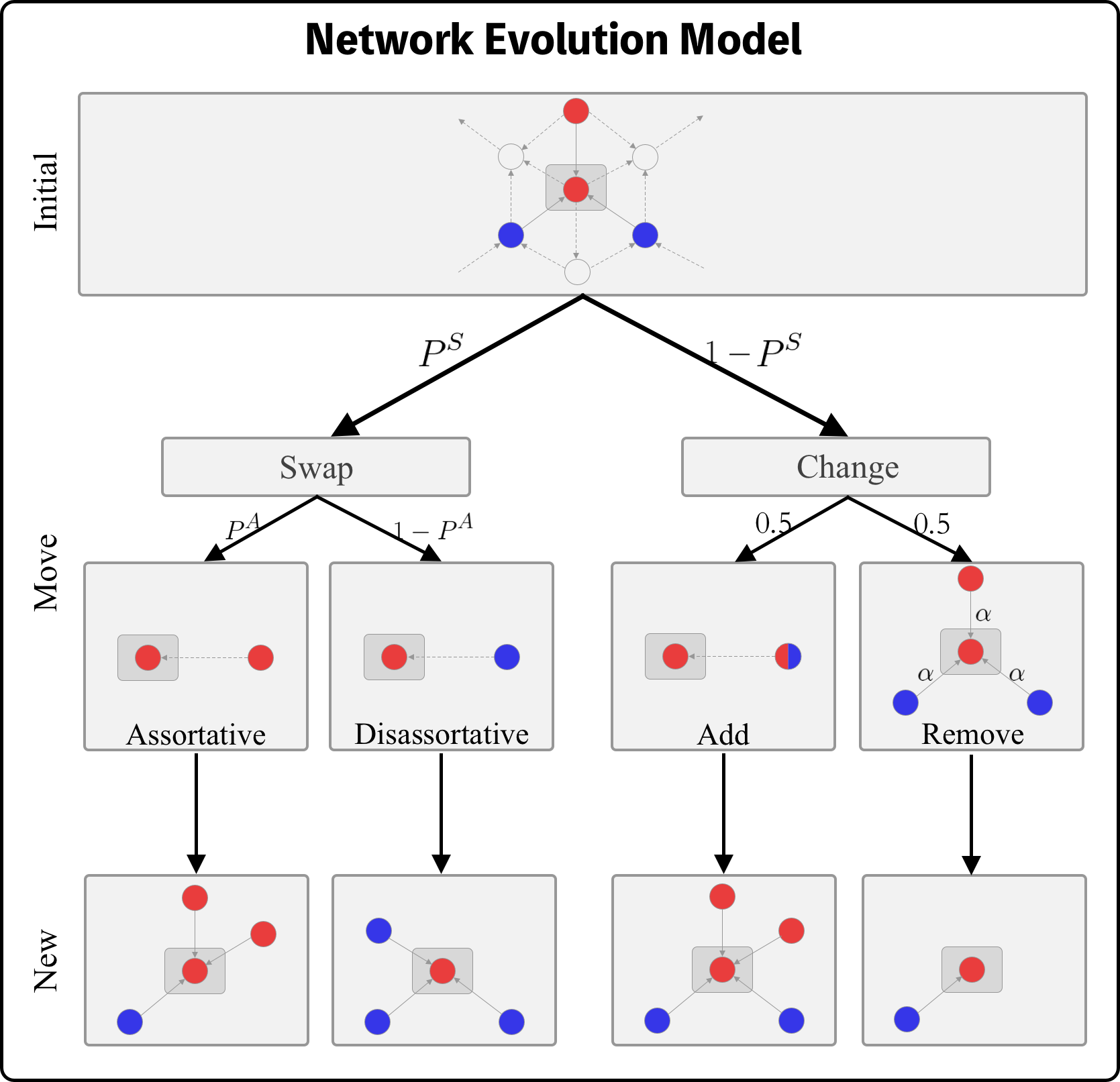}
\centering
\caption{The network evolution model represented as a decision tree. The picture depicts the possible one-step alterations in the connectivity of the network around the randomly chosen focal node (highlighted in the centre of the top panel). The groups of nodes are represented in the picture by colours $g(i)=0$ in red and $g(i)=1$ in blue. The swap move allows users to give preference to edges (influence) from a user of the same group (assortative choice with probability $P^A$) or from the different group (disassortative choice with probability $1-P^A$), while the change move controls the number of incoming edges users in a group will have on average (controlled by parameter $\alpha$). The algorithmic implementation of this model is described in Appendix~\ref{algo:general}.}\label{fig.model}
\end{figure*}

A summary of the evolution is given in Fig.~\ref{fig.model} and a list of parameters in Table ~\ref{table:1}. The choice and application of each move can depend on the group of the involved nodes $g(i)$, but it remains the same for all $t$. We consider $\Delta t = 1/N$ so that $\Delta t \rightarrow 0$ as $N \rightarrow \infty$ and $t \mapsto t+1$ after $N$ steps (i.e., on average each node is picked once).  

\begin{longtable}[!h]{|p{1.6cm}||p{6.1cm}|p{3cm}|}
\hline
\textbf{Parameter} & \textbf{Description} & \textbf{Choice} \\
\hline 
 \multicolumn{3}{|c|}{Primitive Parameters}\\ 
\hline 
  $P^S_g$ & Probability of swap move & $P^S_0 = P^S_1 = P^S$\\
  $P^A_g$ & Probability of assortative move & $0\leq P^A_g \leq 1$\\
  $\alpha_g$ & Probability of removing in-edge & $0\leq \alpha_g \ll 1 $\\
  $N_g$ & Number of nodes & $N_g \gg 1$ \\
  \hline 
 \multicolumn{3}{|c|}{Derived Parameters}\\ 
\hline 
  $\beta_g$ & Proportion of in-edges coming from same group among all in-edges & Eqs.~(\ref{eq:Pr}) and (\ref{eq:PrSolution})\\
  $\langle z \rangle_g$ & Average in-degree & $\langle z \rangle_g = 1/\alpha_g$,  Eq.~(\ref{eq:z*})\\
   $b$ & Ratio between average in-degrees & $b\equiv\langle z \rangle_0/\langle z \rangle_1$\\
    $c$ & Ratio of group sizes & $c \equiv N_0/N_1$\\
\hline
\caption{List of parameters used in our model (4 primitive parameters) and calculations (4 derived parameters). In general, each parameter depends also on the group $g$ of the focal node ($g\in \{0,1\}$). The last column indicates the particular choices and ranges.}\label{table:1}
\end{longtable}
\section{Structure types obtained from the model}

Here we compute the types of structure -- A,D,CP,SB as defined in Fig.~\ref{fig.1} -- obtained for $t\rightarrow \infty$ for different cases of the model introduced above. This is obtained by computing the long-time $t\rightarrow \infty$ density matrix ${\bf \omega}$ in Eq.~(\ref{eq.omegaU}) for the different parameters of our model $P^S_g,P^A_g,\alpha_g,N_g$ and checking if the inequalities~(\ref{eq.inequalities}) defining each community type are satisfied. We are interested in finding the necessary conditions for the appearance of each type and the parameter regime for which a type appears.

Our calculations rely on a key property of our model: the swap move controls assortativity (while the change move is neutral about it) and the change move controls the in-degree (while the swap move preserves it). Accordingly, we start our analysis by computing the average in-degrees $\langle z \rangle_0$ and $\langle z \rangle_1$ as a function of the parameters of the change move and then proceed with the analysis of the effect of the swap move at fixed average in degree. 

\subsection{Change-move determines the average in-degree}

 The primary difference between the remove and add moves is that the former acts on each edge while the latter is a single addition. For large average in-degrees $\langle z \rangle$ and small $\alpha \ll 1$, we can consider the action of this move as providing only a small modification of the in-degree in a single time increment $dt \sim \Delta t = 1/N$. In a continuous and mean-field approximation, the average degree $\langle z \rangle$ of nodes in any of the groups $g$ will change as
\begin{equation}\label{eq.dzdt}
    \frac{d\langle z\rangle}{dt} = -\alpha \langle z \rangle \frac{1}{2} + \frac{1}{2},
\end{equation}
and therefore converge for large $t$ and $N$ towards the fixed point
\begin{equation}\label{eq:z*}
    \frac{d \langle z \rangle }{dt} = 0 \Leftrightarrow\langle z \rangle = \langle z^* \rangle = \frac{1}{\alpha }, 
\end{equation}
where for simplicity we omit the dependence on $g$.

\subsection{General Solution}

In this Section we will compute the expected density matrix $\omega_{rs}$ (at $t\rightarrow \infty$) as a function of the model parameters.
We define $\beta_r(i)$ to be the probability that one of the in-edges of node $i$ is coming from a node $j$ in the same group $g(j)=g(i)=r$. Denoting by $e_{sr}$ the total number of edges from nodes in group $s$ to nodes in group $r$, $\beta_r$ at time $t$ can be computed in a mean-field approximation (i.e., for all $i$ with the same $g(i)=r$) by the macroscopic assortativity of group $g(i)=r$ as
\begin{equation}\label{eq:Pr}
    \beta_{r}(t) = \frac{e_{rr}(t)}{e_{rr}(t)+e_{sr}(t)}
\end{equation}
with $s \neq r$. The denominator of Eq.~(\ref{eq:Pr}) is the total incoming edges for nodes in group $r$.  We obtain the one-time evolution of $\beta_r(t)$ by considering how the number of within-group edges $e_{rr}$ changes in one time step. This is done by computing the probabilities $P^{\pm}$ of increasing (+) and decreasing (-) edges for each of the moves (swap or change) and incorporating the effects into $e_{rr}$
\begin{equation}\label{eq:err1}
  e_{rr}({t+\Delta t}) = e_{rr}({t}) + P^{\text{swap}+} -P^{\text{swap}-}+P^{\text{change}+}-P^{\text{change}-} \langle z \rangle_r ,
\end{equation}
where the last term is multiplied by the average degree $\langle z \rangle$ because the removal of edges in the change move (with probability $\alpha$) applies to each of the existing edges independently. Expressing the probabilities $P^{\pm}$ as a function of the model parameters (see Appendix~\ref{sec.Pr}), and using Eq.~(\ref{eq:Pr}), we can write Eq.~(\ref{eq:err1}) as 
\begin{equation}\label{eq:err2}
e_{rr}({t+\Delta t}) = Be_{rr}({t}) + C,
\end{equation}
with $B = B(t) = 1-P^{S}_r\frac{N_r}{N(e_{rr}(t)+e_{sr}(t))}[P^{A}_r\frac{N_r}{N}+(1-P^{A}_r)(1-\frac{N_r}{N})]-(1-P^S_r)\frac{N_r}{2N(e_{rr}(t)+e_{sr}(t))}$ and $C = C(t) = [P^S_rP^A_r+\frac{1}{2}(1-P^S_r)]\frac{N_r^2}{N^2}$. From Eq.~(\ref{eq:err2}), we obtain the solution as a function of the initial condition at time $t=0$ as 
$$e_{rr}({t}) = B^{Nt}e_{rr}({t=0})+\frac{B^{Nt}-1}{B-1}C.$$ 
Since $B < 1$ for all $t$, $B^{Nt} \rightarrow 0$ and $e_{rr}({t}) \rightarrow \frac{C}{1-B}$ for $t \rightarrow \infty$. In this limit (system at equilibrium), we assume that the total number of edges $e_{rr}({t})+e_{sr}({t})$ is a constant (the expected value at equilibrium) and that the average in-degree can be computed by Eq.~(\ref{eq:z*}). Taking this into account, and introducing the results derived from Eq.~(\ref{eq:err2}) in Eq.~(\ref{eq:Pr}), we obtain the equilibrium probability of an assortative in-edge as 
\begin{equation}\label{eq:PrSolution}
    \beta_r = \frac{P^S_rP^A_r+\frac{1}{2}(1-P^S_r)}{P^S_r(P^A_r+(1-P^A_r)\frac{N_s}{N_r})+(1-P^S_r)\frac{N}{2N_r}}.
\end{equation}

We can now use Eq.~(\ref{eq:PrSolution}) to compute the expected number of edges between groups and therefore the density matrix $\omega_{rs}$. Since in our model we only have two groups, an edge can only be from one group or the other. The probability that an in-edge of node $i$ is from a different group is thus $1-\beta_{r}$. The number of in-edges from nodes in the same group is thus $z_i \beta_r$ and the number of in-edges from the different group is $z_i (1-\beta_r)$. 
Taking this into account, we can compute the number of within group edges $e_{rr}$ and between group edges  $e_{sr}$ as the sum of all edges $\ell \equiv j \mapsto i$ as

\begin{equation}\label{eq:simplee00}
    e_{rr} = \sum_{i=1}^N \delta(g(i)-r) z_i \beta_r = N_r \langle z \rangle_r \beta_r
\end{equation}

\begin{equation}\label{eq:simplee01}
    e_{sr} = \sum_{i=1}^N \delta(g(i)-r) z_i(1- \beta_r) = N_r \langle z\rangle_r (1-\beta_r)
\end{equation}
The density matrix -- defined  in Eq.~(\ref{eq.omegaU}) -- of the network with two groups is then computed as 
\begin{equation}\label{eq:wswap}
  {\bf \omega} =  
    \begin{bmatrix}
\frac{e_{00}}{N_0(N_0-1)} & \frac{e_{01}}{N_0N_1} \\
\frac{e_{10}}{N_0N_1} & \frac{e_{11}}{N_1(N_1-1)} 
\end{bmatrix} 
=
\begin{bmatrix}
\frac{\langle z\rangle_0 \beta_0}{N_0-1} & \frac{\langle z\rangle_1 (1-\beta_1)}{N_0} \\
\frac{\langle z\rangle_0 (1-\beta_0)}{N_1} & \frac{\langle z\rangle_1 \beta_1}{N_1-1}
\end{bmatrix} \\
\approx
\begin{bmatrix}
\frac{\langle z\rangle_0 \beta_0}{N_0} & \frac{\langle z\rangle_1 (1-\beta_1)}{N_0} \\
\frac{\langle z\rangle_0 (1-\beta_0)}{N_1} & \frac{\langle z\rangle_1 \beta_1}{N_1}
\end{bmatrix} 
\end{equation}
where $\beta_0$ and $\beta_1$ are given by Eq.~(\ref{eq:PrSolution}), $\langle z \rangle_0$ and $\langle z \rangle_1$ by Eq.~(\ref{eq:z*}), and the approximate expression is used for large networks ($N_0,N_1 \gg 1$ so that we can drop the subtraction of 1 in the denominator).

\begin{figure*}[h]
\includegraphics[width=0.7\textwidth]{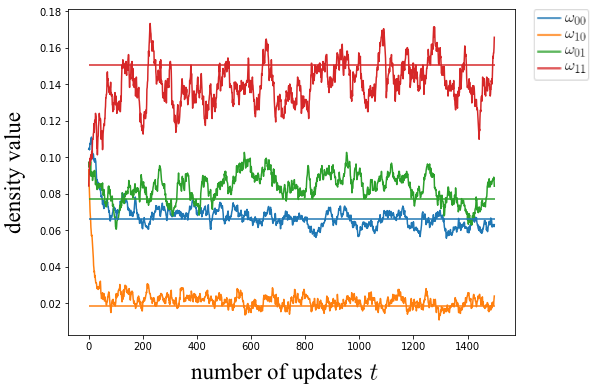}
\centering
\caption{Density of edges $\omega_{rs}$ between nodes in groups $0$ and $1$ as a function of time $t$.  The curves were obtained performing a direct simulation of our model with parameters $P_0^S = 0.7,P_1^S = 0.5, P^A_0 = 0.9,  P^A_1 = 0.8,\alpha_0 = 0.2,\alpha_1 = 0.1,N_0 = 67, N_1 =33$. The horizontal straight line correspond to the expected value for $t\rightarrow \infty$ and $N\rightarrow\infty$ obtained from Eq.~(\ref{eq:wswap}).}\label{fig:Sim}
\end{figure*}

The density matrix in Eq.~(\ref{eq:wswap}), evaluated using Eq.~(\ref{eq:PrSolution}), is the main result of our calculations. In Fig.~\ref{fig:Sim} we show how the results of a direct simulation of our model compares to the prediction from these equations, confirming the agreement of the average density at long times.  Next we discuss the different types of community relationships -- A,CP,D, and SB, obtained from $\omega$ from Eq.~(\ref{eq.inequalities}) -- for different combinations of the model parameters -- $N_0,N_1,P^A_0,P^A_1,P^S_0,P^S_1,\langle z \rangle_0,$ and $\langle z \rangle_1$ --  that appear in the computation of the density matrix in Eq.~(\ref{eq:wswap}).

\subsection{Particular Cases}

\subsubsection{Equal average degrees $\langle z \rangle_0 = \langle z \rangle_1 \Rightarrow$ no SB.}
From the definition~(\ref{eq.inequalities}), the following conditions both hold if SB structure with group $r$ as source and group $s$ as basin is observed
\begin{equation}\label{eq:SB_con1}
    \omega_{rr}>\omega_{rs} \Rightarrow \frac{e_{rr}}{N_r-1} > \frac{e_{rs}}{N_s}
\end{equation}
\begin{equation}\label{eq:SB_con2}
    \omega_{sr}>\omega_{ss} \Rightarrow \frac{e_{sr}}{N_r} > \frac{e_{ss}}{N_s-1}
\end{equation}
with $r,s \in \{0,1\}$ and $r \neq s$.
These two inequalities imply that $\frac{e_{rr}+e_{sr}}{N_r}+\frac{e_{rr}}{N_r(N_r-1)}>\frac{e_{ss}+e_{rs}}{N_s}$. In the limit of sparse networks, $\frac{e_{rr}}{N_r(N_r-1)} \rightarrow 0$ as $N \rightarrow \infty$ and therefore $\frac{e_{rr}+e_{sr}}{N_r}>\frac{e_{ss}+e_{rs}}{N_s} \Rightarrow \langle z \rangle_r > \langle z \rangle_s $. This is in contradiction to our hypothesis of $\langle z \rangle_r = \langle z \rangle_s \Rightarrow \frac{e_{rr}+e_{sr}}{N_r} = \frac{e_{rs}+e_{ss}}{N_s}$. 
This proof by contradiction, which applies not only to our model but to any network partition into two groups, implies that a difference in average degree is essential for the existance of the SB structure (the basin requires a higher in-degree than the source).

\subsubsection{Equal assortativity $P^A_0=P^A_1 \Rightarrow$ no CP.} 
The appearance of CP structure with group $r \in \{0,1\}$ as core and group $s \in \{0,1\}$ ($s \neq r$) as periphery implies the following two mutually conflicting inequalities

\begin{equation}\label{eq:CP_con1}
    \omega_{rr}>\omega_{sr} \Rightarrow \beta_r >\frac{N_r}{N} \Rightarrow P^A_r > \frac{1}{2},
\end{equation}
\begin{equation}\label{eq:CP_con2}
    \omega_{rs}>\omega_{ss} \Rightarrow \beta_s <\frac{N_s}{N} \Rightarrow P^A_s < \frac{1}{2},
\end{equation}
where in the first step we used Eq.~(\ref{eq:wswap}) and in the second step we used Eq.~(\ref{eq:PrSolution}). 
This demonstration shows that groups with equal tendency to establish assortative edges $P^A_0=P^A_1$ could not support a CP structure and the CP structure only appears when one group prefers assortative edges and the other group prefers disassortative edges.

\begin{figure*}[h]
\includegraphics[width=0.85\textwidth]{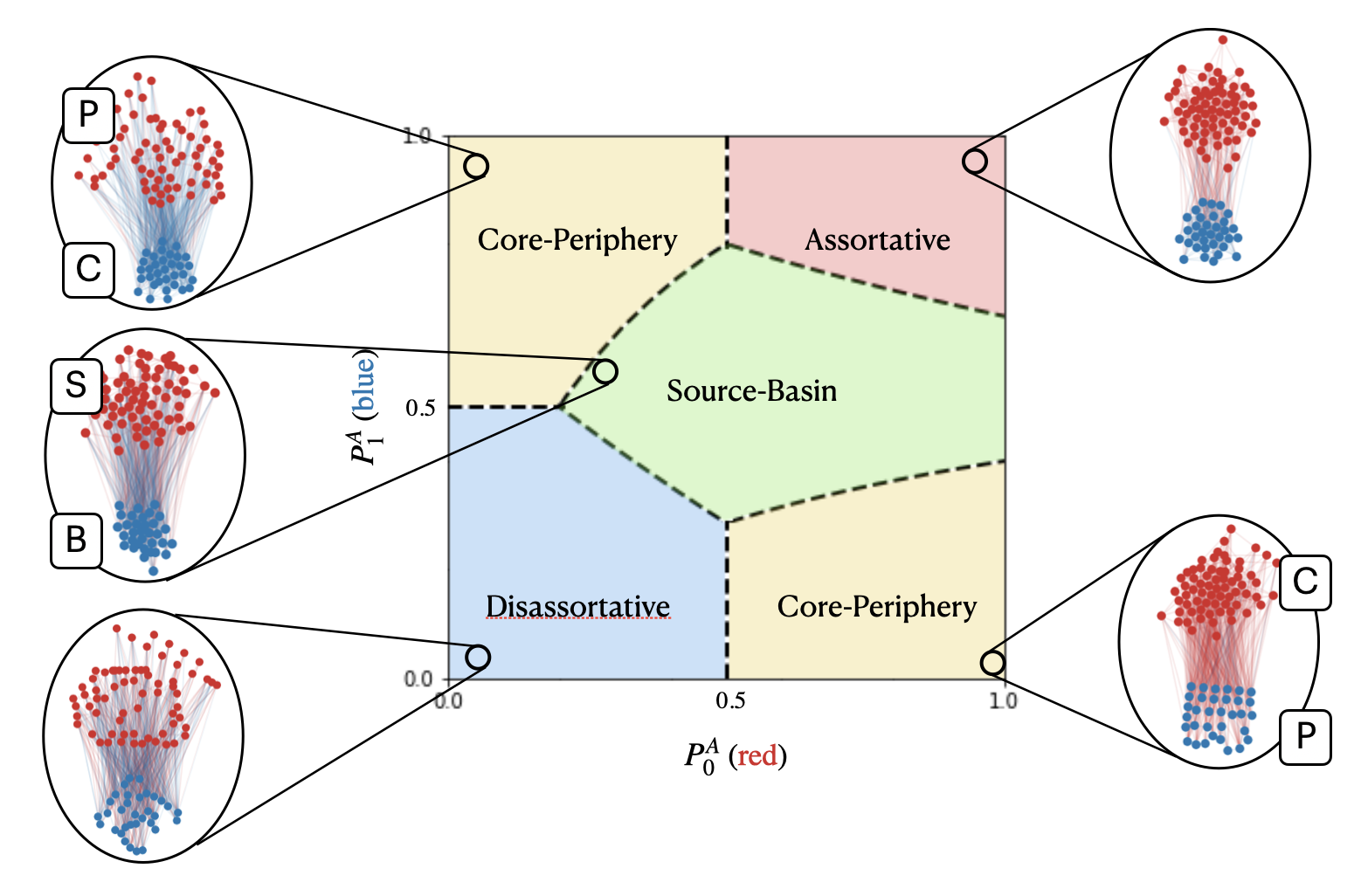}
\centering
\caption{Parameter space for Eq.~\ref{eq:wswap}, which is valid in the limit $P^S\rightarrow 1_-$ with swap and change moves. The group size ratio $c =\frac{N_0}{N_1} =2$ and the average in-degree ratio $b= \frac{\langle z\rangle_0}{\langle z\rangle_1} = 0.5$. Insets: examples of networks obtained at the indicated parameters through direct numerical simulations of our model, plotted using the NetworkX package. The node color indicates the group membership of the node and the edge color indicates information source (same as source node color).}\label{fig:four}
\end{figure*}

\begin{figure*}[h]
\includegraphics[width=0.85\textwidth]{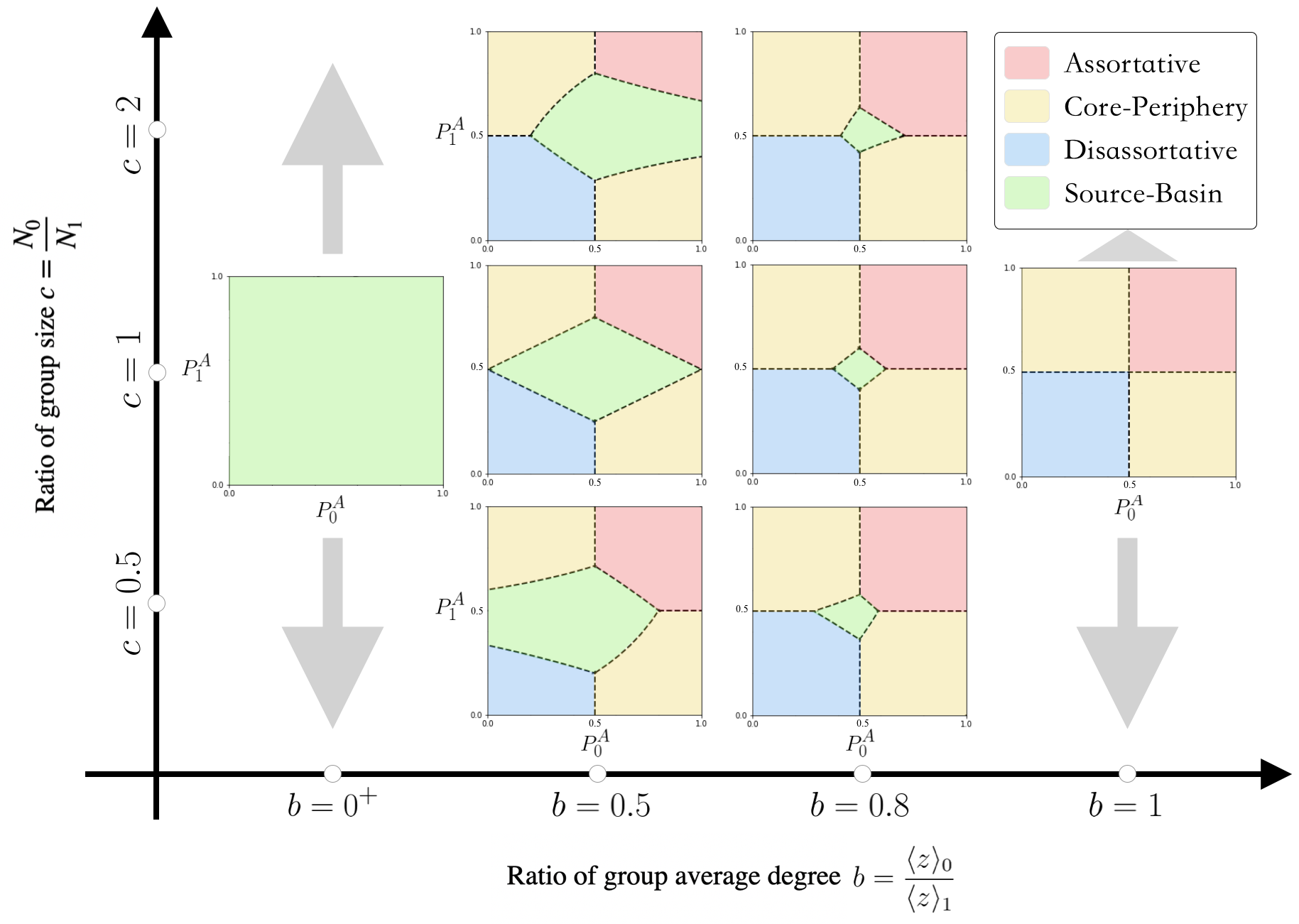}
\centering
\caption{Effect of group-asymmetry in group size $N$ (y axis) and in-degree $\langle z \rangle$ (x axis) in the parameter space of assortative linking shown in Fig.~(\ref{fig:four}). The boundary of all structure types is computed from Eqs.~(\ref{eq:wswap}), (\ref{eq:Pr_sovled}), and (\ref{eq.inequalities}) in the limit of $t\rightarrow \infty$,  $N \rightarrow \infty$, and $P^S\rightarrow 1_-$.}\label{fig:bc}
\end{figure*}

\subsubsection{Swapping regime $P^S\rightarrow 1_-$.} Here we consider the case in which swap moves dominate over change moves. We denote this by $P^S\rightarrow 1_-$ to convey that there are enough change moves to allow the average degree to converge in Eq.~(\ref{eq:z*}) but that otherwise it plays no role. The probability of linking to the same group, $\beta_r$, is obtained by setting $P^S=1$ in Eq.~(\ref{eq:PrSolution}) leading to
\begin{equation}\label{eq:Pr_sovled}
    \beta_{r} = \frac{P_{r}^{A}}{P_{r}^{A}+(1-P_{r}^{A})\frac{N_s}{N_r}}.
\end{equation}
Using this expression in the computation of the density $\omega$ in Eq.~(\ref{eq:wswap}) we obtain a simpler separation of the parameter space into the community types. Fig.~\ref{fig:four} shows that the four structure types appear in the parameter space of $P^A_0$ and $P^A_1$. An A relationship appears when both groups strongly prefer linking to nodes in the same group, while D is formed in the opposite situation. The CP structure appears when one group prefersassortative linking (being the core) while the other group (the periphery) prefers to link to the different group, i.e. it is influenced more by the core group than its own group. SB occurs for intermediate values of $P^A$, most strongly around $P^A_0 = P^A_1 = 0.5$ (no preference in terms of groups).  Fig. \ref{fig:bc} shows the effect of asymmetric group sizes ($N_0\neq N_1$) and average degrees ($\langle z\rangle_0 \neq \langle z\rangle_1$), indicating that the SB region in parameter space grows and is deformed with increasing asymmetry.

\subsubsection{The effect of change moves $P^S<1$.} 
\begin{figure*}[h]
\includegraphics[width=0.85\textwidth]{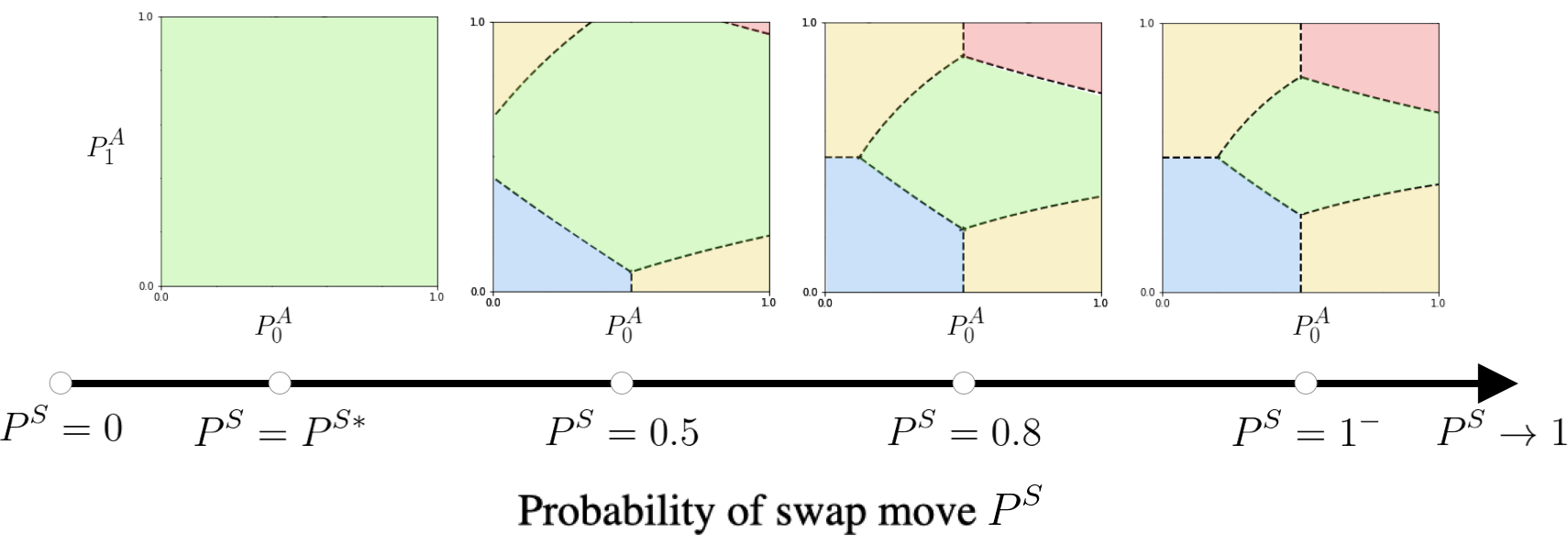}
\centering
\caption{Probability of swapping $P^S$ affects the conditions for the appearance of different community relationships. The results shown in the figure were computed using $c = \frac{N_0}{N_1}=2$, $b = \frac{\langle z\rangle_0^*}{\langle z\rangle_1^*}=\frac{1}{2},$ and $P^S_0 = P^S_1$. The boundary of all structure types is computed with Eqs.~(\ref{eq:wswap}), (\ref{eq:PrSolution}), and (\ref{eq.inequalities}), valid in the limit of $t\rightarrow \infty$ and $N \rightarrow \infty$.}\label{fig:PG}
\end{figure*}

We now consider the effect of change moves in the picture discussed above (i.e., when $P^S<1$). 
For simplicity, we assume the two groups have the same probability of choosing the swap move, i.e. $P^S_0 = P^S_1$. The effect of $P^S$ in reproducing the four community types is shown in Fig. \ref{fig:PG}. As less swap moves are performed ($P^S$ decreases), more in-edges among all edges come from the change move and therefore the assortative probability $P^A$ of the swap move plays a less significant role. The degree difference between the groups becomes increasingly the only important distinction between them. As a result, the area of $SB$ in the $P^A$ parameter space increases with reducing $P^S$.
Surprisingly, there is a finite value $P^{S*}>0$ such that SB is the only community type for any $P^S<P^{S*}$ (see Appendix~\ref{app.psstar} for a computation of $P^{S*}$). An intuitive explanation for these results is that the change move does not take the groups (colours) of nodes into account and therefore it effectively leads to a weakening of the assortative moves ($P^A$ comes closer to $0.5$), which is the SB region in Fig~\ref{fig:four}. 

\section{Conclusions and discussions}

We proposed a simple network-evolution model that generates directed networks with the four different types of relationships between $2$-communities of nodes. Our mean-field calculations of the dependence of the relationship type -- Assortative (A), core-periphery (CP), disassortative (D), and source basin (SB) -- on the model parameters reveal necessary conditions and possible mechanisms behind their creation. In particular, we find that a CP relationship cannot exist if both groups have the same tendency to establish links within their own communities and that a SB relationship cannot exist if the communities have the same average degree. When the two groups have different in-degree, SB appears when the link preference of nodes is not strongly affected by the community assignment of the nodes they link to. Increasing (decreasing) such preference in both groups leads to assortative (disassortative) relationships, while an asymmetric change leads to CP. In general, variations in group sizes and average degree across the groups reduce the parameter range in which A,D, and CP appear and favour the appearance of SB relationships.

Our analysis revealed a close relationship between the difference in in-degree between the communities and the appearance of the Source-Basin community type. A difference in average in-degree is a necessary condition for SB, with the basin being the community with larger average in-degree (i.e., more influenced), and SB appears as the only arrangement in the limit of small $P^S$, which is a limit of a random network with the only difference being the different degree of nodes. These results generalize to the case of directed networks previous analyses on core-periphery in undirected networks, which have shown that CP structure between two groups is highly related with degree difference~\cite{barucca2016disentangling,zhang2015identification} and that no CP structure is found in two groups using a degree-correction method~\cite{kojaku2018core}.
To further clarify the connection between degree and structure type, we consider an alternative definition of a density matrix $\tilde{\omega}$ in which the number of edges $e_{rs}$ is normalized by the product of existing edges $e_re_s$ instead of the possible edges $N_rN_S$. This choice corresponds to a relative connectivity between communities, normalized by their propensity of links, and is similar to the approach used in degree-corrected stochastic block models.  In this case, neither CP nor SB are possible (as shown in Appendix D) in the case of two communities, consistent with previous observations (only A and D are possible).

Our work also suggests different extensions of our model to consider more general cases and more realistic network-evolution processes. For instance, one could consider local searches in the network when re-wiring and also processes in which the group affiliation of nodes changes over time (depending on the neighbours). An important extension would be to consider more than $2$ groups, as this would lift many of the restrictions on the appearance of SB and therefore potentially show a richer picture for  pairwise SB relationships. 
With more than two communities, a pairwise SB relationship can occur because of the excess degree that the two communities acquire due to different interactions with other communities.  An interesting future direction would be to look for signatures of these preferences in real-world network data.

\appendix

\section*{Acknowledgements}
E.G.A. thanks the Max-Planck-Institute for the Physics of Complex Systems (Dresden, Germany) and the Complexity Science Hub (Vienna, Austria) for their hospitality and for enabling insightful discussions.

\section*{Appendices}
\section{Derivation of General Solution}\label{sec.Pr}

In this appendix we show the steps in the demonstration of Eq.~(\ref{eq:err2}). We notice that the within group edges $e_{rr}$ is decided in four possible edge updates:

\begin{itemize}
    \item In swap move, an extra edge adds to $e_{rr}^t$ in the case that an edge $k\mapsto i$ from group $s$ to group $r$ ($g(k)=s,g(i) = r$) rewires to an edge $ j\mapsto i$ coming from group $r$ ($g(j) = r$). This happens when 1) we pick a focal node $i$ in group $r$ with probability $\frac{N_r}{N}$; 2) the node $i$ plays dissassortative game with probability $P^{S}_rP^{A}_r$; 3) we pick an existing edge $k\mapsto i$ that is from different group with probability $1-\beta_r$; and 4) we pick an non-edge $j\mapsto i$ that has $g(j) = r$ with probability $\frac{N_r}{N}$. These $4$ steps are necessary and independent of each other. Therefore, the probability of gaining an edge within group $r$ is calculated as the product of their probabilities as
\begin{equation}\label{eq:addedge}
    P^+(\text{swap}) = \frac{N_r}{N}P^{S}_rP_{r}^{A}(1-\beta_r)\frac{N_r}{N}.
\end{equation}

    \item In swap move, $e_rr^{t}$ decreases one edge in the case that an edge $ k\mapsto i$ from group $r$ to group $r$ ($g(k)=g(i)=r$) rewires to edge $ j\mapsto i$ coming from group $s$ ($g(j)=s$). It happens when 1) we pick a focal node $i$ in group $r$ with probability $\frac{N_r}{N}$; 2) the node $i$ plays disassortative game with probability $P^{S}_r(1-P^{A}_r)$; 3) we pick an existing edge $k\mapsto i$ that is from same group with probability $\beta_r$; and 4) we pick an non-edge $j\mapsto i$ that has $g(j) \neq r$ with probability $\frac{N_s}{N}$. As above, the probability of losing an edge within group $r$ is calculated as

\begin{equation}\label{eq:removeedge}
    P^-(\text{swap}) = \frac{N_r}{N}P^{S}_r(1-P_{r}^{A})\beta_r\frac{N_s}{N}.
\end{equation}

    \item In change move, $e_{rr}^{t}$ increases by 1 when the add move is chosen and a new edge $j\mapsto i$ coming from group $r$ is added at random. It happens when 1) we pick a focal node $i$  in group $r$ with probability $\frac{N_r}{N}$; 2) node $i$ plays add move with probability $(1-P^{S}_r)\frac{1}{2}$; and 3) we pick a non-existing edge to add with approximate probability $\frac{N_r}{N}$. The probability in this case is thus
\begin{equation}\label{eq:addedge3}
    P^+(\text{change}) = (1-P^S_r)\frac{1}{2}\frac{N_r}{N}\frac{N_r}{N}.
\end{equation}
    \item In change move, $e_{rr}^t$ decreases when an edge $ k\mapsto i$ from group $r$ is deleted. It happens when 1) we pick a focal node $i$  in group $r$ with probability $\frac{N_r}{N}$; 2) node $i$ plays remove move with probability $(1-P^{S}_r)\frac{1}{2}$; and 3) we pick every existing incoming edges from group $r$ with probability $\alpha_r \beta_r$. The probability in this case is thus 
\begin{equation}\label{eq:removeedge3}
    P^-(\text{change})=(1-P^S_r)\frac{1}{2}\frac{N_r}{N}\beta_r\alpha_r.
\end{equation}
\end{itemize}

Equation~(\ref{eq:err2}) is obtained introducing Eqs.~(\ref{eq:addedge})-(\ref{eq:removeedge3}) in Eq.~(\ref{eq:err1}) and using the definition of $\beta_r$ in Eq.~(\ref{eq:Pr}).

\section{Varying probability of remove move}\label{app.evolution}

In the evolution model introduced in Sec.~\ref{ssec.evoloution}, we chose the probability of choosing remove or add move to be $\frac{1}{2}$. In this section, we show that this choice is not restricting the cases obtained in our model because varying this probability our density matrix can be made invariant with a proper choice of other parameters. We define the probability of choosing remove move as $P^R$ and the probability of add move as $1-P^R$. This probability affects the average in-degree calculation, with Eq.~(\ref{eq.dzdt}) written as
\begin{equation}\label{eq.dzdt2}
    \frac{d\langle z\rangle}{dt} = -\alpha \langle z \rangle P^R + (1-P^R),
\end{equation}
with the corresponding fixed point solution
\begin{equation}\label{eq:z*2}
    \langle z^* \rangle = \frac{1-P^R}{\alpha P^R}.
\end{equation}
 With different $P^R$, we can choose new $\widetilde{\alpha} =\frac{\alpha P^R}{1-P^R}$ so that Eq.~(\ref{eq:z*}) still holds.

The proportion of assortative edges $\beta_r$ is also affected by $P^R$, with Eq.~(\ref{eq:PrSolution}) generalized to
\begin{equation}\label{eq:Pr3}
    \beta_r = \frac{P^S_rP^A_r+(1-P^S_r)(1-P^R_r)}{P^S_r(P^A_r+(1-P^A_r)\frac{N_s}{N_r})+(1-P^S_r)(1-P^R_r)\frac{N}{N_r}},
\end{equation}
By choosing $\widetilde{P^S_r} = \frac{P^S_r}{2(1-P^S_r)(1-P^R_r)+P^S_r}$, we can also get the same equation as Eq.~(\ref{eq:PrSolution}).

\section{Computation of $P^{S*}$}\label{app.psstar}

Here, we calculate that in the equilibrium of our model, there is a finite value $P^{S*}>0$ such that for any probability of swap move $P^S<P^{S*}$ only SB community types are obtained. We compute $P^{S*}$ by considering the conditions for which no A,D, and CP can be obtained in our model.
We check the conditions of A,D and CP structures from Eqs.~(\ref{eq:PrSolution}) and (\ref{eq.inequalities}) with assumption of $P^S_0 = P^S_1$ and $b<1$.
\begin{enumerate}
    \item Assortative structure can not be obtained if the boundary between A and SB exceeds the probability space, that is, boundary function $P^A_1 > 1$ when $P^A_0 = 1$. The largest probability of swap move that does not show A structure is then $P^A_1(P^A_0=1) = 1$ which we denote as $P^{S*}(A)$ and is computed as the solution (for $x$) of the implicit equation
    \begin{equation}
        \frac{(1-b)(1+x)(1+c-x+cx)}{2x(1+c-x+cx+b(1-c)(1+x))}=1.
    \end{equation}
    
    \item Disassortative structure can not be obtained if the boundary between D and SB exceeds the probability space, that is, boundary function $P^A_1 < 0$ when $P^A_0 = 0$. The largest probability of swap move that does not show D structure is then $P^A_1(P^A_0=0) = 0$ which we denote as $P^{S*}(D)$ and is computed as the solution (for $x$) of the implicit equation
    \begin{equation}
 \frac{bcx^2+\frac{1}{2}x(1-x)(b+2bc-1)+\frac{1}{4}(1-x)^2(bc+b-c-1)}{x^2(1-b+bc)+\frac{1}{2}x(1-x)(c+1+bc+b)}=0.
    \end{equation}
    
    \item Core-periphery structure can not be obtained if the boundary between CP and SB exceeds the probability space. Since we have two cases of CP with 1). group 1 is core and group 0 is periphery and 2). group 0 is core and group 1 is periphery, the conditions of no CP would be $P^A_1 > 1$ when $P^A_0 = 0$ for the first case and $P^A_1 < 0$ when $P^A_0 = 1$ for the second case. The largest probability of swap move that does not show first case of CP is $P^A_1(P^A_0=0) = 1$ which we denote as $P^{S*}(\text{first CP})$ and is computed as the solution (for $x$) of the implicit equation
    \begin{equation}
\frac{(1-bc)x^2+\frac{1}{2}x(1-x)(1+\frac{1}{c
        +1}-\frac{b}{c+1}-bc)+\frac{1}{4}(1-x)^2(c+1-bc-b)}{x^2(1-bc+b)+\frac{1}{2}x(1-x)(c+1-bc^2+b-bc)}=1.
    \end{equation}
    The largest probability of swap move that does not show second case of CP is $P^A_1(P^A_0=1) = 0$ which we denote as $P^{S*}(\text{second CP})$ and is computed as the solution (for $x$) of the implicit equation
    \begin{equation}
\frac{bcx^2+\frac{1}{2}x(1-x)(2bc+b-c)+\frac{1}{4}(1-x)^2(bc-c+b-1)}{x^2(c+bc-b)+\frac{1}{2}x(1-x)(c+1+bc-b)}=0.
    \end{equation}
\end{enumerate}

The value of $P^{S*}$ is thus the minimum value among the above four values
\begin{equation}\label{eq:psstar}
    P^{S*} = \min\{P^{S*}(A),P^{S*}(D),P^{S*}(\text{first CP}),P^{S*}(\text{second CP})\}.
\end{equation}

\section{No SB or CP in modified density matrix}
The density matrix $\omega_{rs}$ in Eq.~(\ref{eq.omegaU}) was defined comparing the number of existing links to the number of all possible links. Here we consider an alternative definition obtained dividing the number of links from group $r$ to $s$ by the product of all existing out-links of group $r$ and in=links of group $s$:
\begin{equation}\label{eq.omega2}
\tilde{\omega} _{rs} = \frac{e_{rs}}{e_r^{\text{out}}e_s^{\text{in}}},
\end{equation}
where $e_r^{\text{out}} = e_{rr}+e_{rs}$ is the total out degrees of group $r$ and $e_s^{\text{in}} = e_{ss}+e_{rs}$ is the total in degrees of group $s$.  This modified density matrix takes degrees of nodes into account and thus removes the effect of degree difference between groups. 

We now show that neither SB nor CP structures can be found if the new density matrix~(\ref{eq.omega2}) is used. We first show that the conditions of SB structure are mutually conflicting. Assuming, without loss of generality, that group $r$ is the source and group $s$ the basin, the conditions for SB in Eq.~(\ref{eq.inequalities}) applied to $\tilde{\omega}_{rs}$ in Eq.~(\ref{eq.omega2}) lead to the following mutually contradicting inequalities
\begin{equation}\label{eq:SB_con3}
    \tilde{\omega}_{rr}>\tilde{\omega}_{rs} \Rightarrow \frac{e_{rr}}{(e_{rr}+e_{sr})(e_{rr}+e_{rs})}>\frac{e_{rs}}{(e_{ss}+e_{rs})(e_{rr}+e_{rs})} \Rightarrow e_{rr}e_{ss}>e_{rs}e_{sr},
\end{equation}
\begin{equation}\label{eq:SB_con4}
    \tilde{\omega}_{sr}>\tilde{\omega}_{ss} \Rightarrow \frac{e_{sr}}{(e_{ss}+e_{sr})(e_{rr}+e_{sr})}>\frac{e_{ss}}{(e_{ss}+e_{sr})(e_{ss}+e_{rs})} \Rightarrow e_{rr}e_{ss}<e_{rs}e_{sr}.
\end{equation}
Similarly, the conditions of CP structure are also mutually conflicting.  Assuming, without loss of generality, that group $r$ is the core and group $s$ the periphery, the conditions for CP in Eq.~(\ref{eq.inequalities}) applied to $\tilde{\omega}_{rs}$ in Eq.~(\ref{eq.omega2}) lead to the following mutually contradicting inequalities
\begin{equation}\label{eq:CP_con3}
    \tilde{\omega}_{rr}>\tilde{\omega}_{sr} \Rightarrow \frac{e_{rr}}{(e_{rr}+e_{sr})(e_{rr}+e_{rs})}>\frac{e_{sr}}{(e_{ss}+e_{sr})(e_{rr}+e_{sr})} \Rightarrow e_{rr}e_{ss}>e_{rs}e_{sr}
\end{equation}
\begin{equation}\label{eq:CP_con4}
    \tilde{\omega}_{rs}>\tilde{\omega}_{ss} \Rightarrow \frac{e_{rs}}{(e_{rr}+e_{rs})(e_{ss}+e_{rs})}>\frac{e_{ss}}{(e_{ss}+e_{sr})(e_{ss}+e_{rs})} \Rightarrow e_{rr}e_{ss}<e_{rs}e_{sr}
\end{equation}

\section{Algorithm}\label{algo:general}
Description of the algorithm used to implement our model of network evolution.

\begin{itemize}
  \item[Initialization:] Erdős-Rényi directed network with $N$ nodes and $2/N<q<1-2/N$ edge creation probability
  \item[1.] Randomly choose one node $i\in[1,N]$. 
  \item[2.] Choose $u \in [0,1]$ uniformly at random and decide update:
    \begin{itemize}
  \item[(Swap)] If $u<P^G(g(i))$:
    \\ Randomly pick $k$ such that $A_{ki}=1$ and $j$ such that $A_{ji} = 0$
    \\ If $g(k)=g(j)$: update ($A_{ki} \leftarrow 1$ and $A_{ji}\leftarrow 0$) or ($A_{ki} \leftarrow 0$ and $A_{ji}\leftarrow 1$) by chance.
    \\ If $g(k)\neq g(j)$: choose $v \in [0,1]$ and:
    \begin{itemize}
     \item[(A)] If $v<P^A(g(i))$ (assortative): 
     \\ if $g(k)=g(i)$: update ($A_{ki} \leftarrow 1$ and $A_{ji}\leftarrow 0$)
     \\ if $g(j)=g(i)$: update ($A_{ki} \leftarrow 0$ and $A_{ji}\leftarrow 1$)
     \item[(D)] If $v\ge P^A(g(i))$ (disassortative): 
     \\ if $g(k)=g(i)$: update ($A_{ki} \leftarrow 0$ and $A_{ji}\leftarrow 1$)
     \\ if $g(j)=g(i)$: update ($A_{ki} \leftarrow 1$ and $A_{ji}\leftarrow 0$)
    \end{itemize}
    \end{itemize}
    \begin{itemize}
        \item[(Change)] If $u\ge P^G(g(i))$:
        \\ Choose $v \in [0,1]$ uniformly at random:
    \begin{itemize}
     \item[(R)] If $v<0.5$ (remove): 
     \\ update $A_{ki} \leftarrow 0$ with probability $\alpha(g(i))$ for all $k$ such that $A_{ki}=1$.
     \item[(A)] If $v \ge 0.5$ (add): 
     \\ update $A_{ki} \leftarrow 1$ for a randomly picked $k$ such that $A_{ki}=0$.
    \end{itemize}
    \end{itemize}
  \item[3.] Update $t\mapsto t+\Delta t$; go to Step 1. 
\end{itemize}

\bibliographystyle{unsrt}

\end{document}